\documentclass{nic-series}
\usepackage{CJKutf8}
\def\msol{{\rm M}_\odot}
\def\zsol{{\rm Z}_\odot}
\def\ltsima{$\; \buildrel < \over \sim \;$}
\def\simlt{\lower.5ex\hbox{\ltsima}}
\def\gtsima{$\; \buildrel > \over \sim \;$}
\def\simgt{\lower.5ex\hbox{\gtsima}}
\begin{document}

\title{Growth of Seed Black Holes in Galactic Nuclei}

\author{
\begin{CJK}{UTF8}{min}
Rainer Spurzem \inst{1,2} \and
 Francesco Rizzuto \inst{3}  \and
 Manuel Arca Sedda \inst{4}  \and
 Albrecht Kamlah \inst{5,1} \and
 Peter Berczik \inst{1,6,7} \and
 Qi Shu (舒琦)\inst{8,2} \and 
 Ataru Tanikawa (谷 川 衝)\inst{9} \and \\
 Thorsten Naab \inst{10} 
 \end{CJK}
 }

\authortoc{R. Spurzem, F. Rizzuto, M. Arca Sedda, A. Kamlah, P. Berczik, et al.}


\institute{Astronomisches Rechen-Institut, Zentrum f\"ur Astronomie,\\
University of Heidelberg, M\"onchhofstrasse 12-14, 69120 Heidelberg, Germany \\ 
\email{spurzem@ari.uni-heidelberg.de}
\and
Kavli Institute for Astronomy and Astrophysics, Peking University, \\
Yiheyuan Lu 5, Haidian Qu, 100871, Beijing, China 
\and
Department of Physics, University of Helsinki, Gustaf Hallstromin katu 2, 00014 Helsinki, Finland
\email{francesco.rizzuto@helsinki.fi}
\and
Dep. of Physics and Astronomy "Galileo Galilei"
University of Padova, Via Marzolo 8, 35131 Padova, Italy 
\email{m.arcasedda@gmail.com}
\and
Max-Planck-Institut
für Astronomie, Königstuhl 17, 69117 Heidelberg, Germany \\
\email{albrechtk@hotmail.de}
\and
Main Astronomical Observatory, National Academy of Sciences of Ukraine, 27 Akademika Zabolotnoho St., 03143 Kyiv, Ukraine
\email{berczik@mao.kiev.ua}
\and
Konkoly Observatory, Research Centre for Astronomy and Earth Sciences, E\"otv\"os Lor\'and Research Network (ELKH), Konkoly Thege Mikl\'os \'ut 15-17, 1121 Budapest, Hungary
\and
Department of Astronomy, School of Physics, Peking University, Yiheyuan Lu 5, Haidian Qu, 100871, Beijing, China
\email{shuqi@pku.edu.cn}
\and
Department of Earth Science and Astronomy, College of Arts and Sciences, The University of Tokyo, 3-8-1 Komaba, Meguro-ku, Tokyo 153-8902, Japan
\email{tanikawa@g.ecc.u-tokyo.ac.jp}
\and
Max-Planck Institute for Astrophysics, Karl-Schwarzschild-Str. 1, 85741 Garching, Germany \email{naab@MPA-Garching.MPG.DE}
}

\maketitle

\begin{abstracts}
The evolution of dense star clusters is followed by direct high-accuracy N-body simulation. The problem is to first order a gravitational N-body problem, but stars evolve due to astrophysics and the more massive ones form black holes or neutron stars as compact remnants at the end of their life. After including updates of stellar evolution of massive stars and for the relativistic treatment of black hole binaries we find the growth of intermediate mass black holes and we show that in star clusters binary black hole mergers in the so-called pair creation supernova (PSN) gap occur easily. Such black hole mergers have been recently observed by the LIGO-Virgo-KAGRA (LVK) collaboration, a network of ground based gravitational wave detectors.
\end{abstracts}

\section{Introduction}

It is one of the grand challenges of theoretical astrophysics to understand the dynamics of dense star clusters, both in their form as galactic globular clusters orbiting in the Milky Way halo as well as nuclear star clusters, surrounding the central supermassive black holes (SMBH) in our Galaxy and in other galaxies. High precision dynamical simulations of star clusters use direct orbit integration under the influence of (in principle) all other stars allowing precise modeling of diffusive transport processes of mass, energy, and angular momentum in the star cluster. This is important to understand relaxation processes in the system. The physical and astrophysical challenge is not only the gravitational million-body problem, but also the presence of a large number of very tight binaries and stellar evolution with black holes, neutron stars and white dwarfs forming. This becomes a strong multi-scale problem, with orbital time scales of days coupling to overall crossing times of the cluster of million years and ages of billion years.

The evolution of dense star clusters is not only governed by the aging of their stellar populations and simple Newtonian dynamics. The stellar densities become so high that stars can interact and collide, stellar evolution and binary stars change the dynamical evolution, black holes can accumulate in their centers and merge with relativistic effects becoming important. Recent high-resolution imaging has revealed even more complex structural properties with respect to stellar populations, binary fractions and compact objects as well as – the still controversial – existence of intermediate mass black holes in clusters of intermediate mass. Dense star clusters therefore are the ideal laboratory for the concomitant study of stellar evolution and Newtonian as well as relativistic dynamics. Last but not least black holes forming and evolving in dense star clusters are one of the prominent sources of gravitational waves (GW) across all frequency windows. 

\section{Current Astrophysical Updates}
\label{sec1}

\subsection{Stellar Evolution of Massive Stars}
\label{subsec1}
Before the first LIGO-Virgo-KAGRA (LVK) gravitational wave (GW) detection, many theoretical models of stellar evolution predicted stellar black holes (BHs) masses to be lower than 30 $\msol$. These models remained unchallenged for several years because all stellar BHs observed at the time had masses $\simlt\!20\,\msol$ (Ziółkowski 2008; Özel et al. 2010). Surprisingly, the first LVK detection, GW150914, revealed components more massive than 30 $\msol$ (Abbott et al. 2016). Such masses had been predicted by stellar evolution models at low metallicity introducing a dependence between stellar winds mass loss and metallicity (see Woosley et al. 2002; Vink et al. 2001, and references therein).

This highlights the importance of up-to-date stellar evolution models for the correct interpretation and prediction of GW events. An accurate theory for the evolution of massive stars is particularly important to predict the mass distribution of stellar BHs at their formation. For this, precise models of stellar winds and a correct description of the last stages of
the stellar evolution before the collapse are required. At the onset of stellar collapse, stars with sufficiently large helium cores undergo a phase of electron-positron pair production that in turn leads to one or more violent explosions. Depending on the initial mass of the core,
the star can experience pulsation pair-instability supernovae (PPSN) getting partially destroyed or it can experience the more violent pair-instability supernovae (PSN) and is destroyed completely \cite{Fowler1964,Woosley2007,Woosley2017}. Due to (P)PSN, isolated massive stars are not supposed to collapse into BHs in the mass range of approximately $50\!-\!130\,\msol$. This gap in the stellar
BH mass distribution is known as the (P)PSN mass gap. The mass
limits of this gap are affected by various uncertainties and therefore they depend on the details of the stellar evolution adopted. In this study, the assumed mass gap is $45\!-\!195\,\msol$.

\begin{figure}[t]
\begin{center}

\includegraphics[width=12cm]{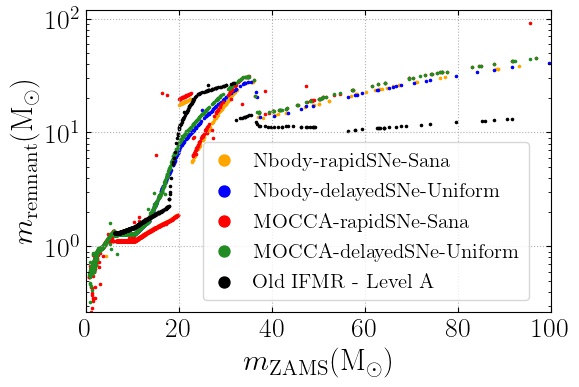}
\caption{\label{ifmr-mocca-nbody}
 Initial-Final mass relation (IFMR) for the escaping compact objects of the MOCCA and Nbody6++GPU (Nbody) simulations. The keys refer to the Nbody-delayedSNe-Uniform, Nbody-rapidSNe-Sana, MOCCA-delayedSNe-Uniform, and MOCCA-rapidSNe-Sana simulations, respectively. They differ by the use of the code (Nbody or MOCCA), and the remnant mass prescription for the stellar mass black holes up to approximately the pair-instability mass gap (see Fryer et al. 2012). The black points show BH masses from another $N$-body simulation with Level A parameters\cite{Belczynski2002} (Plot taken from \citen{Kamlah2022}). 
}
\end{center}
\end{figure}

Most of current updates of stellar evolution have been published for NBODY7\cite{Banerjee2020} and for NBODY6++GPU\cite{Kamlah2022}. The major issues are summarized here:
\begin{itemize}
\item[(i)] New stellar wind models following \citen{Belczynski2010}, which in turn follow the wind mass-loss rates given by \citen{Vink2001}. With these models, BHs masses that originate from single stars depend strongly on the metallicity. For instance, a $100
\msol$ isolated main-sequence star would leave a 15 $\msol$ BH at so-
lar metallicity ($Z = 0.02$). At very low metallicity ($Z = 0.0002$),
however, it can form a BH of about $60 \msol$, in the absence of pair-instability models.
\item[(ii)] Pair-instability supernova and pulsation pair-instability supernova models (according to \citen{Belczynski2016} incorporated in the remnant formation and supernovae models as described in \citen{Fryer2012}. Stars with helium core with masses $> 40 \msol$ undergo a violent phase of mass loss. For helium cores in the range between $60\! - \! 135/,\msol$ the star is completely destroyed.
\item[(iii)] New prescription of BHs and NSs natal kick velocities that explicitly depend on the fallback fraction\cite{Banerjee2020}.
\item[(iv)] A model for electron capture supernovae (ECSN) that produces neutron stars with low-velocity kicks that are therefore likely retained in medium-size star clusters\cite{Podsiadlowski2004,Gessner2018}.
\end{itemize}

Fig.~\ref{ifmr-mocca-nbody} shows an example of how important the new updates of stellar evolution are. It shows the initial-final mass relation (IFMR), which tells what is the mass of a stellar evolution remnant as a function of the initial mass of a star. On the y-axis all objects more massive than a few solar masses are black holes; the figure compares the old IFMR with two variants of the new one (delayed and rapid supernovae explosions), for our direct N-body simulations as well as for an approximate Monte Carlo model (MOCCA)\cite{Giersz2015,Giersz2019}.

\subsection{Preparation for PopIII stars}

Fitting formulae have been worked out for evolution tracks of massive stars with $ \simlt M\simlt 160\,\msol$ under extreme metal poor (EMP) environments for $\log(Z/\zsol) = -2$, -4, -5, -6, and -8, where $\msol$ and $\zsol$ are the solar mass and metallicity, respectively\cite{Tanikawa2020}. They are based on reference stellar models, newly obtained by simulating the time evolutions of EMP stars. The fitting formulae take into account stars ending as blue supergiant (BSG) stars, and stars skipping the Hertzsprung
gap (HG) phases and blue loops, which are characteristic of massive EMP stars. Here stars may remain BSG stars when they finish their core Helium burning (CHeB) phase. The fitting formulae are in good agreement with the stellar evolution models, and are now used for our NBODY6++GPU code within the SSE/BSE packages; they are also used for other codes such as PeTar and NBODY7 (see discussion in Section~\ref{sec3}). The algorithms should be useful to generate theoretical predictions for black holes and black hole mergers under EMP environments. The work with NBODY6++GPU is currently in progress on the Juwels-Booster system (Kamlah, Tanikawa, et al., in preparation). 

\subsection{Relativistic Dynamics of Compact Objects}

The gravitational energy loss and resulting merger of compact
objects is computed following the orbit-averaged approach\cite{Peters1963}, allowing for the final coalescence if the orbit shrinking time due to gravitational wave emission becomes shorter than an orbital time. We have added the following further updates, which affect the formation and evolution of black holes\cite{Rizzuto2021}:

\begin{itemize}
    \item[(i)] For collisions between a compact remnant and a main sequence star or red giant a free parameter $f_c$ is introduced, which describes the mass loss from the system in the process. The previous NBODY6 versions used only $f_c = 1$, i.e. no mass loss in the process.
    \item[(ii)] Simultaneous treatment of classical tidal interactions (Roche lobe overflow) and Post-Newtonian orbit-averaged orbit shrinking due to gravitational wave emission has been made possible. Both are treated technically in a similar way, and can now be switched on together.
    \item[(iii)] Strongly bound binaries of two compact objects, which are subject to Post-Newtonian relativistic energy loss are prevented from unperturbed two-body integration, and defined as a new type of binary in the code.
\end{itemize}

\begin{figure}[t]
\begin{center}
\includegraphics[width=12cm]{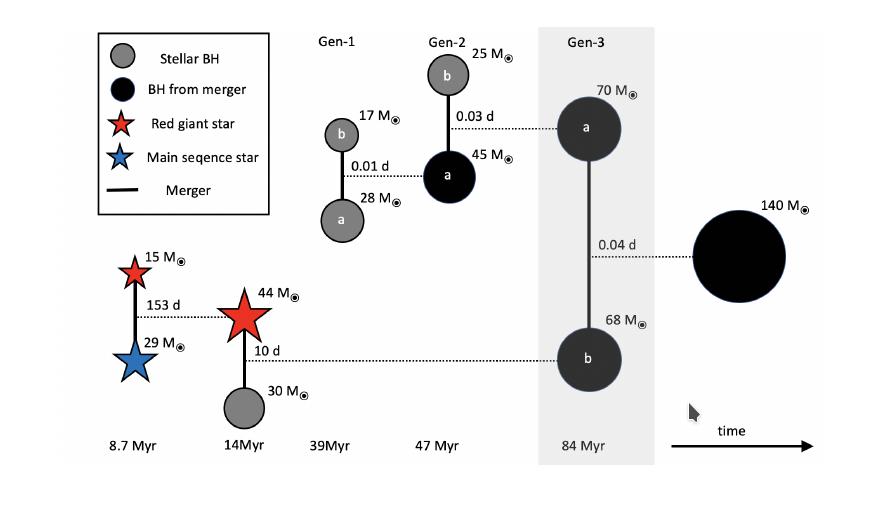}
\caption{\label{mass-gap-merger}
Visualization of the formation path towards a ”mass gap” merger (grey region, third generation) of two black holes with mass $70\msol$ and $68\msol$ developed in one of our N-body simulations. The more massive BH (third generation) grew by two preceding mergers involving black holes (first and second generation). The lower mass BH (third generation) was created in a stellar merger of a red giant with a main sequence star followed by the collision with a stellar mass BH. The masses of the components (in solar masses) and the orbital periods (in days) are indicated at the respective times of the merger (black horizontal lines) after the start of the simulation (Figure from \citen{ArcaSedda2021}).
}
\end{center}
\end{figure}

Fig.~\ref{mass-gap-merger} shows an example, and illustrates, how an intermediate mass black hole is formed through several steps, involving collisions of two massive stars, a collision between a massive star and a black hole and relativistic mergers of black hole binaries. The procedure and parameters described above will affect rates and timescales of the black hole formation process.

\subsection{Relativistic Recoils at Coalescence and Spins}

Another important physical process that has not yet been used for the published papers so far is relativistic recoil for compact object coalescences. Its absence might artificially enhance the probability of forming massive black holes\cite{Banerjee2021,Banerjee2022}. We have implemented now the relativistic recoils following \citen{Morawski2018}, and currently tests are ongoing on the Juwels-Booster system (Arca Sedda et al., in prep.).

Regarding spins (have been neglected so far, and are important to determine proper values of recoil at coalescence) we have now included in our simulation code a new spin treatment\cite{Belczynski2020}, which provides three alternative models for spin evolution of stars, with variable efficiency of angular momentum transport, as well as further options to choose different spin configurations for single and binary black holes (Arca Sedda et al., in prep.). The one with highest efficiency by \citen{Fuller2019} delivers at the end point of massive star evolution black holes with low spins, consistent with current LVK observations. 

Last, but not least the ultimate goal is to use a full Post-Newtonian dynamics inside the regularized binary motion. In such approach relativistic periastron precession and energy loss due to gravitational radiation can be followed at each point of the orbit using a generalized quasi-classical equation of motion\cite{Kupi2006}. Recently also the inclusion of spin dynamics and spin-spin and spin-orbit interactions are included into the Post-Newtonian approach in our codes\cite{Sobolenko2021}. Our cited papers describe the method to integrate this into our codes; any reader interested in the relativistic theory for Post-Newtonian dynamics please refer to papers cited therein.

\section{Results}

\subsection{Intermediate Mass Black Hole Formation}
Young dense massive star clusters are a promising environment for the formation of intermediate mass black holes (IMBHs) through collisions (of massive stars) and coalescences (of smaller mass black holes). We have published a set of 80 simulations carried out with NBODY6++GPU using 10 different initial conditions, and shown that an IMBH can form in some cases with – so far – up to $350\,\msol$. We simulated compact star clusters with $1.1\cdot 10^5$ particles ($\sim 7 \cdot 10^4\,\msol$, core density of $\sim 10^5\,\msol {\rm pc}^{-3}$) with a resolved stellar population with 10\% initial (primordial) hard binaries, and find that very massive stars with masses up to $\sim 400\,\msol$ grow rapidly by binary exchange and three-body scattering events with main sequences stars in hard binaries. From them IMBHs with masses up to $350\,\msol$ form on timescales of order 15 Myr; the final mass depends critically on an unknown parameter describing how much mass is accreted, if a black hole collides with a main sequence star (see description above of $f_c$ parameter).

\begin{figure}[t]
\begin{center}

\includegraphics[width=12cm]{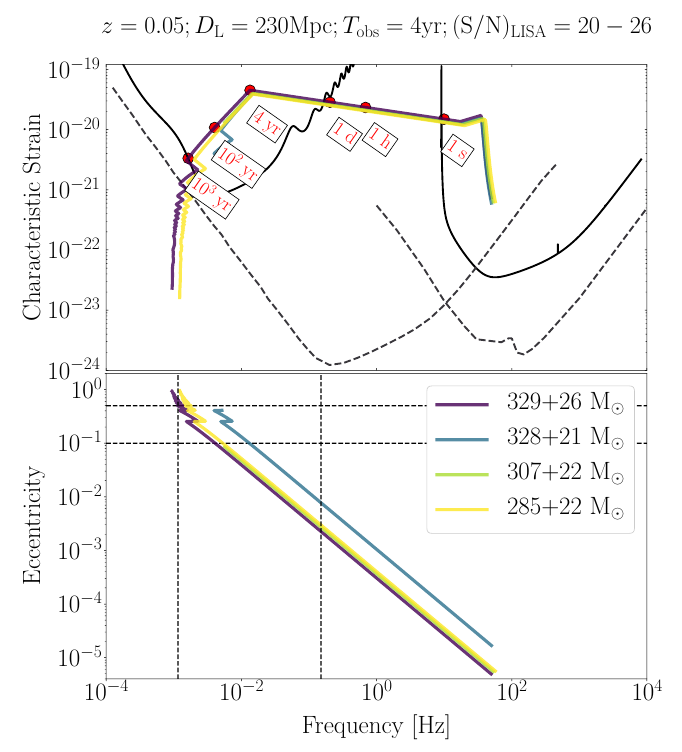}
\caption{\label{GW-strain-evolution}
Top panel: GW strain evolution as a function of the frequency for the mergers beyond the mass-gap (coloured lines).
Simulated tracks are overlapped to the sensitivity curves of LISA and LIGO-Virgo-KAGRA (LVK, solid black lines), and DECIGO
and Einstein Telescope (dashed black lines). The white boxes indicate the time to merger for the heaviest merger.
Bottom panel: eccentricity evolution as a function of the frequency. The vertical lines identify the moment in which the
IMBH-BH mergers enter and exit the LISA sensitivity window, whereas horizontal lines identify the eccentricity values
$e = 0.1, 0.5$. All the mergers are assumed to happen at a redshift $z = 0.05$, corresponding to a luminosity distance
$D_L = 230$ Mpc. The figure title report the typical signal-to-noise ratio (S/N) $\,\sim\! 20\! - \! 26$, assuming for LISA a $T_{\rm obs} = 4$ yr long mission. Figure from \citen{ArcaSedda2021}.
}
\end{center}
\end{figure}

It is the first time that such IMBH formation process has been shown in a direct N-body simulation of that quality and particle resolution. This process was qualitatively predicted from Monte Carlo MOCCA simulations\cite{Giersz2015,Giersz2019}. After formation, the IMBHs can experience occasional mergers with stellar mass black holes in intermediate mass-ratio inspiral events on a 100 Myr timescale. For more details please compare \citen{Rizzuto2021}. 

\subsection{Black Hole Merger in the Forbidden Zone}

The LIGO-Virgo-KAGRA Collaboration (LVC) discovered GW190521, a gravitational wave (GW) source associated with the merger between two black holes (BHs) with masses of 66 and $>\!85\,\msol$. GW190521 represents the first BH binary merger with a primary mass falling in the PSN mass gap (see explanation in Section~\ref{subsec1}) and leaving behind a $\sim 150\,\msol$ remnant. So far, the LVC has reported the discovery of four further mergers having a total mass $>\!100\,\msol$, i.e., in the intermediate-mass black hole (IMBH) mass range. In our simulations we discover the development of a GW190521-like system as the result of a third-generation merger, and furthermore four IMBH-BH mergers with total mass $(300-350)\,\msol$. We show that these IMBH-BH mergers are low-frequency GW sources detectable with LISA and Decihertz Interferometer Gravitational wave Observatory (DECIGO) out to redshift $z = 0.01-0.1$ and $z > 100$, and we discuss how their detection could help unraveling IMBH natal spins. For the GW190521 test case, we show that the third-generation merger remnant has a spin and effective spin parameter that matches the 90\% credible interval measured for GW190521 better than a simpler double merger and comparable to a single merger. Due to GW recoil kicks, we show that retaining the products of these mergers require birth sites with escape velocities $\,\simeq\! 50\!-\!100$ km/s, values typically attained in galactic nuclei and massive clusters with steep density profiles. So, this is an explanation why observed black hole mergers do occur in the PSN mass gap – they are the result of several generation mergers in star clusters\cite{ArcaSedda2021,Rizzuto2022}.

\begin{figure}[t]
\begin{center}

\includegraphics[width=12cm]{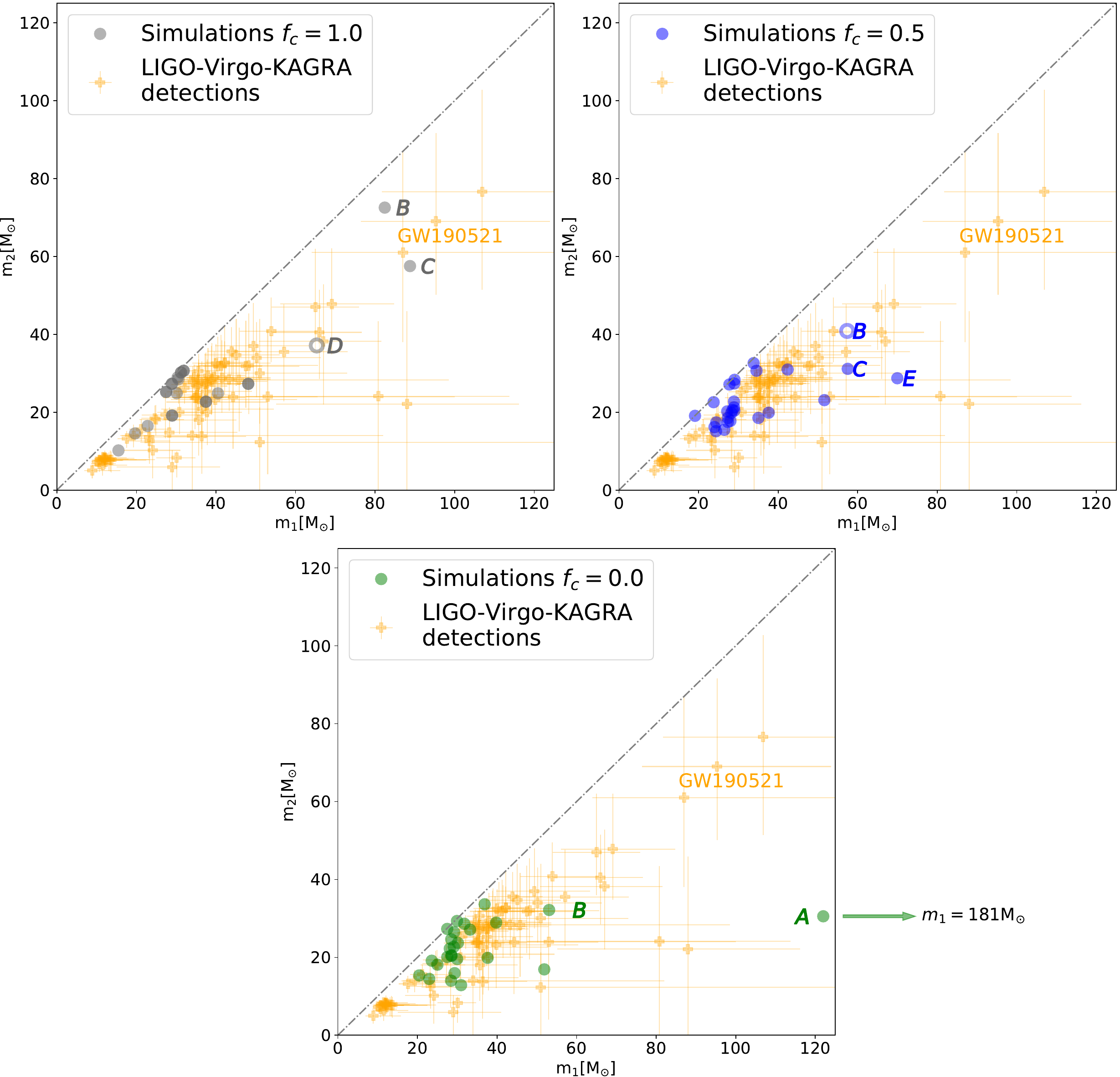}
\caption{\label{bh_bh_collisions}
The panels show the primary ($m_1$) and secondary ($m_2$) masses of all BH mergers in the simulations for an accretion fraction of $f_{\rm c}=1.0$ (left, grey circles), $f_{\rm c}=0.5,$ (center, blue circles) and $f_{\rm c}=0.0,$ (bottom, green circles). The currently available LIGO-Virgo-KAGRA gravitational wave detections including error bars are indicated in orange. BH merger events that might be excluded due to gravitational recoil kicks are indicated with open circles. In general, the simulated events cover a similar parameter space as all currently available observations. The $f_{\rm c}=1.0,$ simulations provide two possible formation paths for GW190521. Path C is a second-generation event and has a low probability due to a first-generation BH merger. Path B is more likely as the event itself is a first-generation BH merger. One of the $f_{\rm c}=0.0,$ realizations generated an intermediate-mass ratio inspiral of two black holes with 31 and 181 $\msol$), respectively, as shown in the bottom panel (Figure from \citen{Rizzuto2022}).
}
\end{center}
\end{figure}

We have presented more direct N-body simulations, carried out with NBODY6++GPU, of young and compact low-metallicity (Z = 0.0002) star clusters\cite{Rizzuto2022}. Early on, after tens of Myrs, every simulated cluster hosts several black hole merger events which nearly cover the complete mass range of primary and secondary black hole masses for current LIGO-Virgo-KAGRA gravitational wave detections. The importance of gravitational recoil is estimated statistically during post-processing analysis. We presented possible formation paths of massive black holes above the assumed lower PSN mass-gap limit ($45\,\msol$) into the intermediate mass black hole (IMBH) regime ($>100\,\msol$) which include collisions of stars, black holes and the direct collapse of stellar merger remnants with low core masses. Fig.~\ref{GW-strain-evolution} shows how the black hole binaries found in our simulations would show up in the detection sensitivity diagrams of current ground and future space based gravitational wave detectors. Finally, Fig.~\ref{bh_bh_collisions} compares in a statistical way current LIGO-Virgo-KAGRA detections of binary black hole mergers with a collection of such mergers from our simulations, using in three different panels a different mass loss factor $f_c$ for star-black hole collisions. While we need more observational and simulation data to improve the statistical quality it can be seen already that such comparisons would allow to constrain the physics of black hole star collisions. In a similar way in the future modelling the spin of black holes could and comparing with observations could tell us something about initial and final spins of black holes (which is quite difficult to observe directly via gravitational waves, at least currently).

\section{Initial Models and Codes}
\label{sec3} 

\subsection{Initial Models}
The simulations take into account full stellar evolution as well as the formation and evolution of binary stars. We set the number of primordial binaries typically to be 5 or 10 \% of the total number of systems (these are persistent hard binaries with binding energy higher than the r.m.s. random kinetic energy of a star; a much larger number of soft binaries present at the time of formation of the star cluster is being disrupted early on).

\subsection{NBODY6++GPU}
NBODY6++GPU\footnote{Link to repositories: tarballs/git service at \\ \url{https://zenodo.org/record/6511341/files/Nbody6\%2B\%2BGPU-Jan2022.tgz} \\ \url{https://github.com/kaiwu-astro/Nbody6PPGPU-beijing}}
is a high-precision direct N-body simulation
code based on the earlier N-body codes NBODY1-6\cite{Aarseth1999}
and NBODY6++\cite{Spurzem1999}. It uses for time integration Taylor
series up to $4^{\rm th}$ order; due to the Hermite scheme it can be based on two time points only. This together with the hierarchically blocked variable time step scheme allows an efficient parallelization of the code for massively parallel supercomputers (since NBODY6++);
gravitational forces between particles are offloaded to graphics processing units (GPUs), used for high-performance general purpose
computing (NBODY6++GPU\cite{Wang2015}). The parallelisation is achieved via MPI and OpenMP on the top level, distributing work within a group of particles due for time integration, and efficient parallel use of GPU cores at the base level (every MPI process
using a GPU), for computing the gravitational forces between particles. The GPU implementation in NBODY6++GPU provides a
significant performance improvement, especially for the long-range
(regular) gravitational forces (see \citen{Nitadori2012,Wang2015}). Benchmarks and profiling are published in detail in \citen{Wang2016,Huang2016}.

Recipes to simulate single, binary, and multiple objects stellar evolution are based on the SSE and BSE programs by Hurley\cite{Hurley2000,Hurley2002,Hurley2005} (see also code references \citen{Hurley2013a,Hurley2013b}). It includes rapid tidal circularization for binaries with small pericenters and tidal captures\cite{Mardling2001}. The integrator fully resolves orbits and dynamical evolution of binaries, even during phases of mass loss or when one of the two stars undergoes a supernova explosion. The binary orbit is adjusted to the corresponding loss of mass, energy and angular momentum with appropriate time stepping; in case of a supernova explosion it is always ensuring that the remnant and its companion leave the explosion with the corrected orbital positions and velocities. Further recent improvements are metallicity dependent winds, delayed or rapid white dwarf kicks\cite{Fryer2012}, electron-capture and pair instability supernova events. The reader interested in more details please refer to our papers \citen{Banerjee2020,Kamlah2022}, which summarize the code updates and give all necessary further references. 

\subsection{New developments}

\begin{figure}[t]
\begin{center}

\includegraphics[width=6cm]{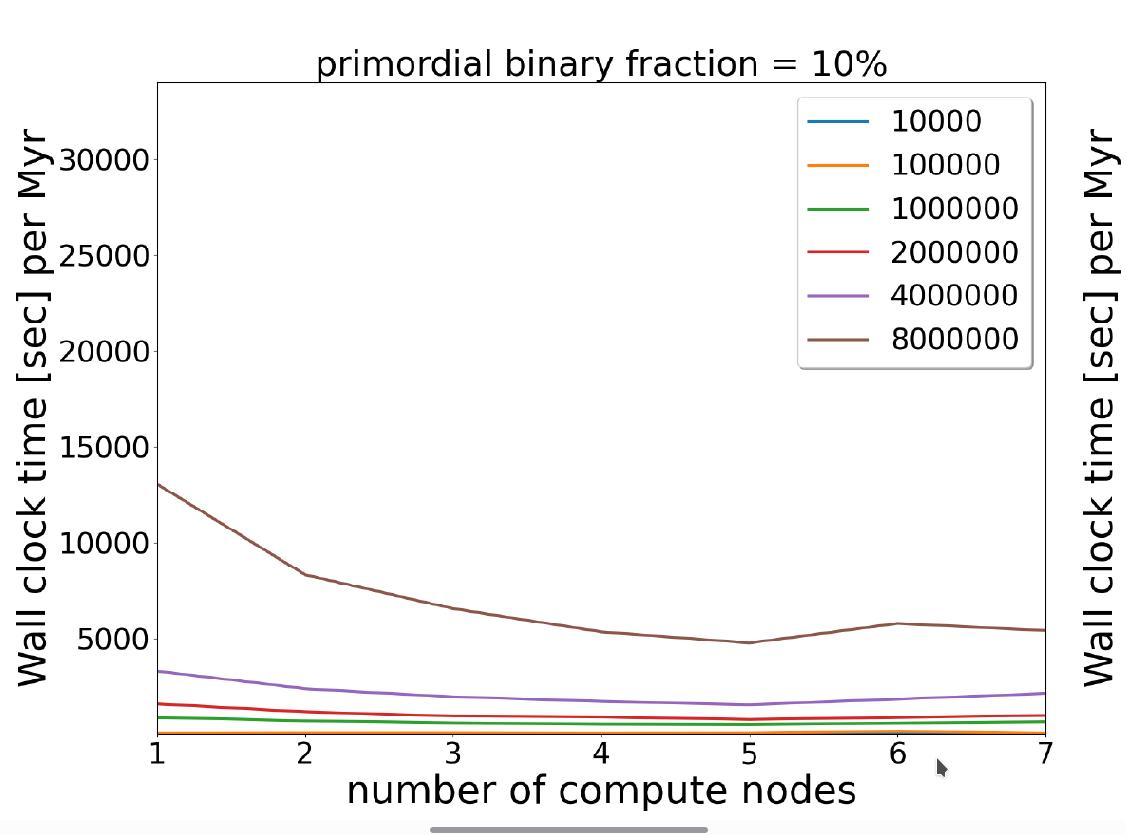}
\includegraphics[width=6cm]{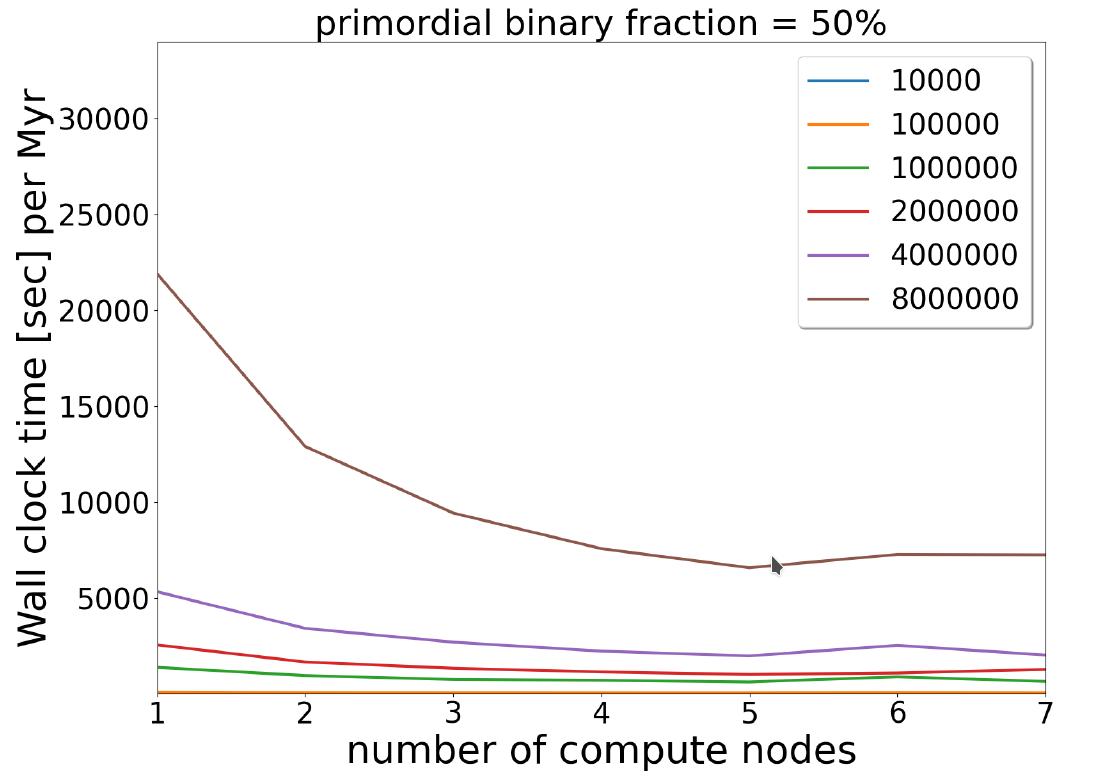}
\caption{\label{petar-benchmark}
Strong scaling of the novel PeTar code, showing wall clock times obtained on the Juwels-Booster as a function of number of compute nodes.
}
\end{center}
\end{figure}

A completely new code, called PeTaR\cite{Wang2020} has been introduced. This code is written in a coherent, modular way. It is supposed to contain all the physics of NBODY6++GPU, but since it is a new code a lot of testing and comparison is still required. Algorithmically it has two advantages over NBODY6++GPU currently, one is the efficient parallelization of hard binaries, and the other is the replacement of distant regular gravitational forces from particles by a TREE based scheme, which makes it possible to take into account also small perturbations on binaries from distant objects without too much computational cost. It is possible to upgrade NBODY6++GPU accordingly; we expect that both codes will co-exist for foreseeable time. Nevertheless we have intensively tested PeTaR, see Fig.~\ref{petar-benchmark}and may use it also in the next computing period in this project.

A novel hierarchical $4^{\rm th}$ fourth-order forward symplectic integrator and its numerical implementation has been shown in a new GPU-accelerated direct-summation N-body code named FROST\cite{Rantala2021}. The new integrator also uses an innovative MSTAR chain scheme\cite{Rantala2020} instead of the classical and algorithmic chains by \citen{Mikkola1998}. The integrator claims to be especially suitable for simulations with a large dynamical range due to its hierarchical nature, and for direct-summation N-body simulations beyond $N = 10^6$ particles on systems with several hundred and more GPUs. In that respect it is very similar to the classic $4^{\rm th}$ order Hermite codes such as $\varphi$GRAPE\cite{Harfst2007,Berczik2013} or HiGPU\cite{Capuzzo2013}. These codes, lacking the Ahmad-Cohen neighbour scheme, can easily used hundreds of GPUs, because they compute full long-range gravitational forces even for the smallest time steps. So, they gain parallelism by introducing order N unnecessary computations, which can be avoided by using NBODY6++GPU. In its AC neighbour scheme full force calculations (done on the GPUs) are only done in order ten times larger time intervals than the smalles steps. For the smallest steps only order 50-200 neighbour particle forces are required for force calculation, and this is done efficiently by using OpenMP on the multi-core host CPU\cite{Wang2015,Wang2016}. Therefore, NBODY6++GPU often appears to be not to scale well to large GPU numbers, because it already obtains physical results comparable to $\varphi$GRAPE or HiGPU (with many GPUs) by using only few GPUs and OpenMP with CPU cores efficiently. 

It should be noted, however, that the new code PeTar uses a TREE scheme for distant forces, which in principle can be competitive with the AC scheme. A full quantitative profiling analysis of these issues is still missing. However, Fig.~\ref{petar-benchmark} shows first benchmarks done by our team using the PeTar code, which are very promising and open up the path to $10^7$ particles in direct N-body simulation, especially with large binary fractions (up to 50\%, a region in which NBODY6++GPU currently lags behind).

\section*{Acknowledgements}
The authors gratefully acknowledge the Gauss Centre for Supercomputing e.V. (www.gauss-centre.eu) for funding this project by providing computing time through the John von Neumann Institute for Computing (NIC) on the GCS Supercomputer JUWELS (2021) at Jülich Supercomputing Centre (JSC). 
MAS is a Marie Curie Skłodowska Fellow and Alexander von Humboldt Fellow. 
AK is a fellow of the International Max
Planck Research School for Astronomy and Cosmic Physics at the
University of Heidelberg (IMPRS-HD). PB and RS acknowledge the support of the Volkswagen Foundation under the Trilateral Partnerships
626 grant No. 97778 and the Strategic Priority Research Program (Pilot B) Multi-wavelength gravitational wave universe of Chinese Academy of Sciences (No. XDB23040100). PB acknowledges support
from Volkswagen Foundation under the special stipend No. 9B870 (2022), from a President's International Fellowship for Visiting Scientists of Chinese Academy of Sciences, from the Science Committee of the Ministry of Education and Science of the Republic of 
Kazakhstan (Grants No. AP08856184 and AP08856149), from the National Academy of Sciences of Ukraine under the Main Astronomical Observatory GPU computing cluster project No.~13.2021.MM, and by
the special program of the National Research Foundation of the Ukraine ‘Leading and Young Scientists Research Support’ - “Astrophysical Relativistic Galactic Objects (ARGO): life cycle of active nucleus”, No. 2020.02/0346. We thank Sambaran Banerjee, Mirek Giersz, Jarrod Hurley, Arek Hypki, Nadine Neumayer, Long Wang, Kai Wu, Roberto Capuzzo-Dolcetta, Andreas Just, M.B.N. (Thijs) Kouwenhoven, Xiaoying Pang for helpful discussions, collaboration, and hospitality during visits. MAS acknowledges funding from the European Union’s Horizon 2020 research and innovation programme under the Marie Skłodowska-Curie grant agreement No. 101025436 (project GRACE-BH, PI Manuel Arca Sedda).


\end{document}